# Structural and magnetic properties of La$_{2/3}$D$_{1/3}$MnO$_3$ (D = Ca, Sr, Ba) manganites prepared by Ball milling

Sami Mahmood[1,a], Haneen Badran[1], Eman Al-Hwaitat[1], Ibrahim Bsoul[2], Rola bqaeen[3]

[1]Department of Physics, The University of Jordan, Amman 11942, Jordan, [2]Department of Physic, [2]Al al-Bayt University, Mafraq 13040, Jordan,[3] Cell thereby center, The University of Jordan, Amman 11942, Jordan

[a*]s.mahmood@ju.edu.jo, (corresponding author),

## Abstract

In this research, we report on the synthesis and characterization of La$_{0.67}$Ca$_{0.33}$MnO$_3$ (LCMO), La$_{0.67}$Ba$_{0.33}$MnO$_3$ (LBMO), and La$_{0.67}$Sr$_{0.33}$MnO$_3$ (LSMO) perovskites. Precursor powders for the perovskite samples were prepared using ball mailing technique. The resulting powder was pressed into disks which were subsequently sintered at 1000º C for 2h. The structural characteristics of the prepared samples were investigated using X-ray diffraction (XRD) and scanning electron microscope (SEM), and the magnetic properties were investigated using vibrating sample magnetometry (VSM). XRD pattern of LCMO sample revealed a pure perovskite phase with *Pnma* space group, while the pattern of LSMO sample revealed a pure perovskite with $R\bar{3}c$ space group. XRD pattern of LBMO sample, however, demonstrated the presence of a major perovskite phase with $R\bar{3}c$ symmetry, together with a secondary BaMnO$_3$ phase. This secondary phase disappeared upon sintering LBMO at a higher temperature of 1100º C. Isothermal magnetic measurements and thermomagnetic curves revealed that LCMO was paramagnetic at room temperature. However, LSMO exhibited soft ferromagnetic behavior at room temperature, with $T_c$ = 380 K and $M_s$ = 50.9 emu/g. Also, LBMO sample sintered at 1000º C exhibited soft ferromagnetic behavior at room temperature, with Curie temperature $T_c$ = 343 K and a rather low saturation magnetization of $M_s$ = 30.7 emu/g. The higher sintering temperature of LBMO compound resulted in a significant increase of the saturation magnetization to 50.8 emu/g for the sample sintered at 1100º C.

# 1. Introduction

Doped perovskite manganites with the general formula $T_{1-x}D_xMnO_3$, where T is a trivalent lanthanide cation and D is a divalent cation, have been the focus of many studies due to their interesting structural, magnetic, transport and electronic structure properties [1]. Perovskites are of vital interest to a wide community of materials scientists, not only because of the fascinating phenomena and properties exhibited (such as colossal magneto-resistance (CMR), metal-insulator transition, charge ordering (CO), phase separation, and magneto-caloric effect (MCE) [2-5]), but also due to their potential technological applications in magnetic recording [6] and gas sensors [7-9].

LaMnO$_3$ compound is an antiferromagnetic insulator below Néel temperature ($T_N = 143$ K) [10]. The partial substitution of La$^{3+}$ ions by a divalent ion D$^{2+}$ (such as Ba$^{2+}$, Sr$^{2+}$, Ca$^{2+}$, Pb$^{2+}$) induces the conversion of some of the Mn$^{3+}$ ions into tetravalent (Mn$^{4+}$) ionic state, where the fraction of the tetravalent ions is equal to the fraction of the divalent ions according to the formula $La_{1-x}D_xMn^{3+}_{1-x}Mn^{4+}_xO_3$. This is responsible for the development of a ferromagnetic state in the manganites, and the wealth of fascinating structural, magnetic, magnetocaloric, magnetotransport, and other phenomena observed in these compounds. Traditionally, double exchange (DE) interactions between Mn$^{3+}$ and Mn$^{4+}$ ions were proposed to explain the magnetic and magnetotransport behavior in these compounds. Theoretical studies, however, have demonstrated that DE alone does not explain the magnetotransport behavior in La$_{1-x}$D$_x$MnO$_3$ [11, 12]. The structural characteristics, magnetic properties, and Curie temperature of the compound are critically sensitive to the type and concentration of the divalent ion [1]. The Ca$^{2+}$ doped manganites (LCMO) were found to exhibit ferromagnetic behavior below room temperature, which is a disadvantage in using these materials in devices operating at room temperature. Also, significant fluctuations in the Curie temperature of this compound were found in the literature, which cannot be associated with Ca$^{2+}$ concentration ($x$) in the compound, or the method of preparation in a systematic manner. Examples of reported values of Curie temperatures for La$_{1-x}$Ca$_x$MnO$_3$ compounds are listed in Table 1. It is important to remember that variations from $x = 0.3$ to $x = 0.33$ are negligible [13].

Table 1: Curie temperature of $La_{1-x}Ca_xMnO_3$ compounds prepared by different methods

| $x$ | $T_c$ (K) | Preparation method (Heat treatment) | Reference |
|---|---|---|---|
| 0.20 | 256 | Sol−gel method (1000º C) | [14] |
| 0.20 | 236 | Sol−gel method (1000º C) | [15] |
| 0.25 | 224 | Sol−gel method (1000º C, 1400º C) | [16] |
| 0.30 | 220 | Solid State Reaction (1200º C) | [17] |
| 0.30 | 255 | Solid State Reaction (1250º C, 1400º C) | [18] |
| 0.33 | 260 | Solid State Reaction (1100º C, 1200º C, 1300º C) | [13] |
| 0.33 | 260.4 | Solid State Reaction (800 º C, 1300º C) | [19] |
| 0.50 | 230 | Pyrophoric method (1000º C) | [20] |

Upon substituting $La^{3+}$ by $Pb^{2+}$ (LPMO), $Ba^{2+}$ (LBMO), or $Sr^{2+}$ (LSMO), the Curie temperature was found to increase to room temperature and above. At $x = 0.3$, Curie temperature for LPMO, LBMO, and LSMO were found to be 300 K, 330 K, and 370 K, respectively [13, 17]. Also, Curie temperature for $La_{0.67}Ba_{0.33}MnO_3$ was reported to be 335.1 K [19] and 332 K [21], and Curie temperature for $La_{0.7}Sr_{0.3}MnO_3$ reported by different research groups was found to be 342.6 K [19], 354 K [22], 369 K [10], and 378.1 K [23]. In addition, a systematic study of the effect of partial substitution of $La^{3+}$ by $Pb^{2+}$ revealed transformation of the compound to a ferromagnetic state, with Curie temperature increasing with the level of substitution from 235 K for $x = 0.1$ to 360 K for $x = 0.4$, and improvement of the magnetocaloric effect [24]. Others, however, reported a rather low Curie temperature of 150 K for $La_{0.67}Sr_{0.33}MnO_3$, and weak magnetization with almost linear behavior in the field range > 2000 Oe [25]. Further, the substitution of $La^{3+}$ by mixtures of divalent ions was used to tailor the Curie temperature and properties of the perovskites for application in specific working temperature regimes [3, 26-30]. However, unsystematic variations of the Curie temperature of manganites produced by different methods at different research laboratories were observed for compounds with nominally identical compositions [1].

A large amount of research work was dedicated to understand the rich structural and magnetic phase diagram exhibited by these compounds [10, 13, 31, 32]. The room-temperature structural results of Urushibara et al. [10] indicated that $La_{1-x}Sr_xMnO_3$ perovskite exhibited orthorhombic *Pbnm* symmetry in the range $0 < x < 0.175$, with lattice parameters: 5.54 Å > $a$ > 5.51 Å, 5.75 Å > $b$ > 5.51 Å, and 7.69 Å < $c$ < 7.80 Å. At higher Sr doping up to $x = 0.4$, the structure changes to rhombohedral symmetry with $R\bar{3}c$ space group. In another study, the lattice parameters of the compound (LSMO, $x = 0.33$) were found to be: $a = b = 5.4879$ Å, $c = 13.3622$ Å [33]. These parameters are somewhat lower than the values observed by Zhang et al. ($a = b = 5.5212$ Å, $c = 13.3797$ Å) [19], and are in better agreement with the lattice parameters of $a = b = 5.5039$ Å, $c = 13.3553$ Å reported for (LSMO, $x = 0.3$) [34]. On the other hand, $La_{0.67}Ca_{0.33}MnO_3$ (LCMO, $x = 0.33$) was characterized by orthorhombic *Pnma* structure with lattice parameters: $a = 5.435$ Å, $b = 7.691$ Å, $c = 5.436$ Å [35]. Zhang et al., however, reported an orthorhombic *Pbnm* structure with lattice parameters of $a = 5.4744$ Å, $b = 5.4601$ Å, $c = 7.7147$ Å for (LCMO, $x = 0.33$) [19]. Also, LCMO, $x = 0.3$ exhibited orthorhombic *Pbnm* symmetry with lattice parameters $a = 5.462$ Å, $b = 5.478$ Å, $c = 7.720$ Å [18]. In addition, the structural characterization of $La_{2/3}(Ca_{1-x}Sr_x)_{1/3}MnO_3$ indicated that the structure of the compound with $x = 0$ is orthorhombic with *Pbnm* space group, and that Sr doping levels above 0.05 ($x \geq 0.15$) transforms the structure to rhombohedral with $R\bar{3}c$ space group [32]. At this point, it is worth mentioning that *Pbnm* space group is equivalent to *Pnma* with a different choice of axes, where $a$, $b$, $c$ in *Pbnm* become $c$, $a$, $b$ in *Pnma* [1]. Further, a comprehensive neutron diffraction study of $A_{0.7}A'_{0.3}MnO_3$ (A = Pr, La, La-Pr; A' = Ca, Sr, Ba, Ca-Sr, Ba-Sr) perovskites was conducted by Radaelli et al. [36]. Room temperature neutron diffraction results indicated that all compounds with $Ca^{2+}$ fraction $\geq 0.17$ are orthorhombic with *Pnma* space group and lattice parameters: 5.4585 Å < $a$ < 5.4690 Å, 7.6749 Å < $b$ < 7.7292 Å, and 5.4308 Å < $c$ < 5.5045 Å. However, all LSMO, LBMO, and LBSMO compounds revealed rhombohedral $R\bar{3}c$ symmetry with lattice parameters: $a = b = 5.5022$ Å – 5.5378 Å and $c = 13.3291$ Å – 13.5011 Å.

Several experimental techniques were used for the preparation of precursor powders of magnetic oxides [37-42]. Ball milling is one of the important commonly used methods due to its simplicity and flexibility of controlling the processing parameters in producing a wide range of materials [43-45]. In this work, we report on the ball-milling synthesis, and structural and magnetic studies of $La_{0.67}D_{0.33}MnO_3$ (D = Ca, Sr, Ba) perovskites. This work focuses on the effects of the type of divalent ion on the magnetic properties and magnetocrystalline anisotropy in these compounds.

## 2. Experimental

### 2.1. Materials preparation

In this study, homogeneous precursor powder mixtures of $La_{0.67}D_{0.33}MnO_3$ manganites (D = Ca, Sr, and Ba) were prepared by ball milling stoichiometric amounts of the starting powders. The starting powders were high-purity $La_2O_3$, $DCO_3$ and $MnO$ with molar ratios of 0.335:0.330:1.000. The powders were mixed and milled in a planetary ball mill (Pulverisette-7) with powder to ball mass ratio of 1:12. The milling was carried out for 16 h at a rotational speed 250 rpm. Disc-shape pellets were made by pressing parts of the powder mixtures in a stainless steel die under the pressure of 5 T. Then the pellets were sintered in a box oven at 1000° C for 2 h using a temperature rate of increase of 10° C/min. Since LBMO was not a pure phase as revealed by XRD data, a sample of this compound was sintered at 1100° C in order to investigate the effect of sintering temperature on the structural properties of this compound.

### 2.2. Materials characterization

The structural characteristics of the synthesized samples were examined by X-ray diffraction (XRD). XRD patterns were collected in the angular range $20° \leq 2\theta \leq 70°$ with a step of 0.01° and scan speed of 0.5°/min. The particle size and morphology of the prepared materials was examined by scanning electron microscopy (SEM). The magnetic properties of the samples were investigated using a conventional vibrating sample magnetometry (VSM) operating at an applied magnetic field up to 10 kOe.

## 3. Result and discussion

### 3.1. XRD measurements

Fig. 1 shows that the XRD patterns of $La_{0.67}Sr_{0.33}MnO_3$ (LSMO) and $La_{0.67}Ca_{0.33}MnO_3$ (LCMO) samples revealed the presence of a single perovskite phase with no detectable secondary phases. However, XRD pattern of $La_{0.67}Ba_{0.33}MnO_3$ sintered at 1000° C (LBMO 1000), showed a perovskite phase, and a secondary phase as revealed by the additional reflections at $2\theta$ = 25.8° and 31.356°, 41.1°, and 55.9° (Fig. 2). These reflections are consistent with those of the standard pattern for $BaMnO_3$ (JCPDS: 00-026-0168). The appearance of a secondary phase in (LBMO 1000) sample is an indication of incomplete crystallization of the perovskite phase at a sintering temperature of 1000° C, and the attainment of a pure perovskite LBMO phase may require modification of the synthesis route and heat treatment. Accordingly, another sample of LBMO was sintered at 1100° C (LBMO 1100) and studied to examine the effect of sintering temperature on the structural and magnetic properties of this compound.

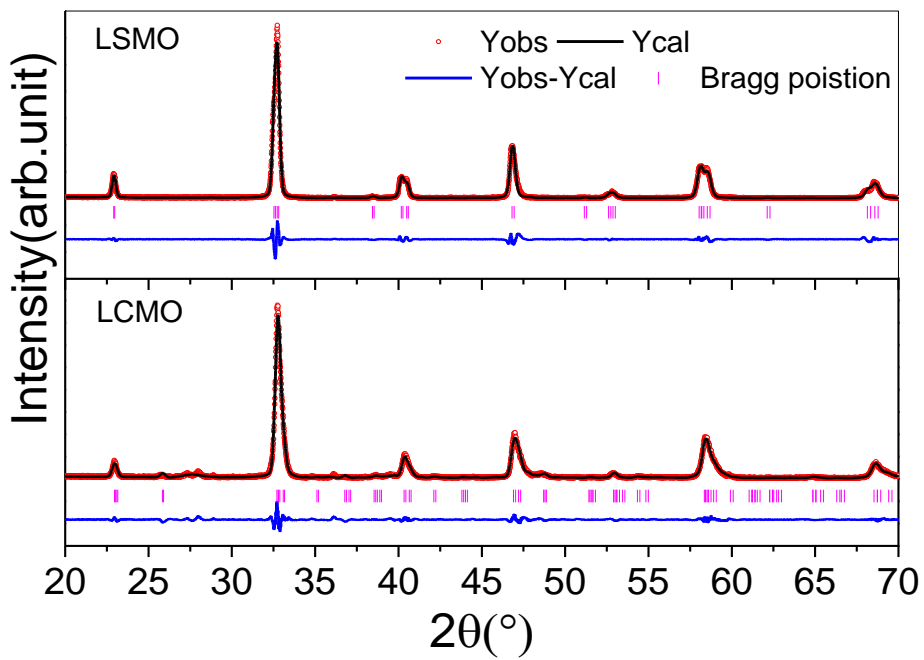

**Fig. 1:** Rietveld plots of XRD data of $La_{0.67}Ca_{0.33}MnO_3$ (LCMO) and $La_{0.67}Sr_{0.33}MnO_3$ (LSMO) compounds.

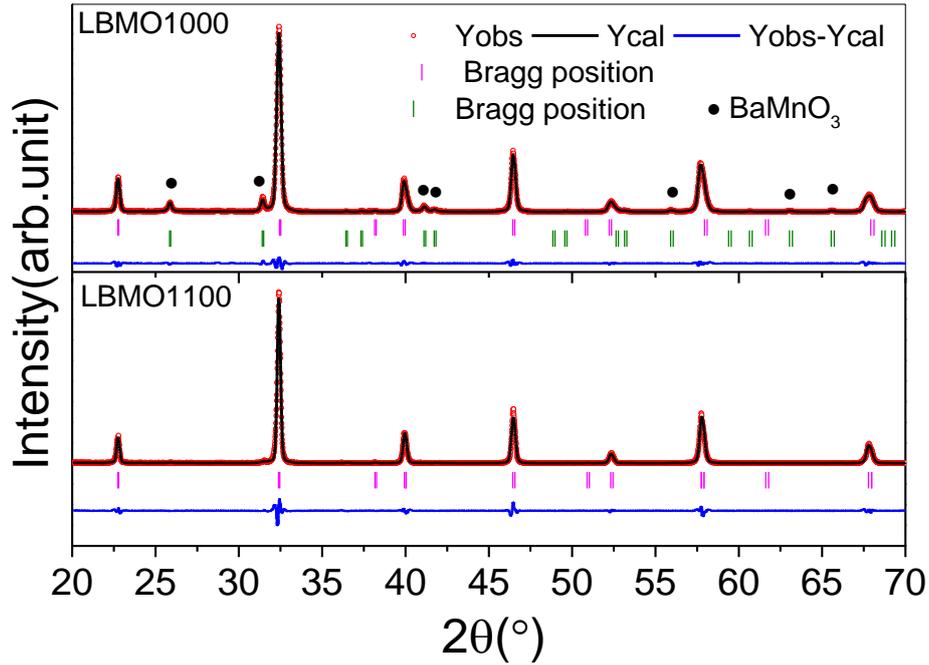

**Fig. 2:** Rietveld plots of XRD data of $La_{0.67}Ba_{0.33}MnO_3$ samples sintered at 1000° C (LBMO 1000), and at 1100° C (LBMO 1100).

Comparison of the XRD patterns of (LBMO 1000) and (LBMO 1100) samples in Fig. 2 revealed significant improvement of the crystallization of the perovskite phase at 1100° C. This improvement was demonstrated by the disappearance of the reflections corresponding to the secondary $BaMnO_3$ phase, and the reduction of the width of the structural peaks of LBMO perovskite phase, indicating that LBMO perovskite phase requires higher formation temperature in comparison with LSMO and LCMO.

Rietveld analysis of the diffraction patterns of all samples was carried out to investigate the microstructural characteristics, and obtain the refined structural parameters of each sample. The refined lattice parameters $a$, $b$, and $c$, as well as the cell volume $V$ and X-ray density $\rho_x$ of the samples are listed in Table 1. The analysis revealed that the LCMO sample consisted of a single perovskite phase with orthorhombic *Pnma* symmetry. This structural symmetry and observed lattice parameters are consistent with previously reported structural results of LCMO [19, 32, 35]. However, both (LBMO 1100) and LSMO samples revealed the presence of a single perovskite phase with rhombohedral $R\bar{3}c$ symmetry. The lattice parameters of LSMO are in good agreement with previously reported values based on XRD and neutron diffraction data for LSMO, $x = 0.3, 0.33$ [33, 34, 36]. Also, the lattice parameters of LBMO samples are in very good agreement with the values ($a$ = 5.5378 Å, $c$ = 13.5011 Å) derived from neutron diffraction data [36], and the values ($a$ = 5.5320 Å, $c$ = 13.5273 Å) [19] and ($a$ = 5.5191 Å, $c$ = 13.5509 Å) [46] derived from XRD data for (LBMO; $x$ = 0.33). Further, the observed refined lattice parameters for (LBMO; $x$ = 0.33) are in agreement with the values ($a$ = 5.5336 Å, $c$ = 13.4838 Å) derived from XRD data for (LBMO; $x$ = 0.3) [47]. The structural symmetry of our LBMO samples, however, is different from the reported cubic $Pm$-$3m$ symmetry with lattice parameter $a$ = 3.9075 Å [48], even though the reported X-ray density of 6.719 g/cm$^3$ is almost equal to our observed value. Although the cell volume increases from LCMO to LBMO, the X-ray density increases progressively, which can be associated with the progressive increase of the molecular mass of the perovskite.

**Table 1:** Refined lattice parameters and cell volume, x-ray density, Mn−O bond length and Mn−O−Mn bond angle of the perovskite structure in the samples.

| Sample | LCMO | LSMO | LBMO1000 | LBMO1100 |
|---|---|---|---|---|
| Space group | $Pnma$ | $R\bar{3}c$ | $R\bar{3}c$ | $R\bar{3}c$ |
| $a$ (Å) | 5.42 | 5.50 | 5.54 | 5.53 |
| $b$ (Å) | 7.75 | 5.50 | 5.54 | 5.53 |
| $c$ (Å) | 5.48 | 13.35 | 13.49 | 13.53 |
| $V$(Å$^3$) | 230 | 350 | 358 | 358 |
| $\rho_x$ (g/cm$^3$) | 6.07 | 6.41 | 6.72 | 6.72 |
| Mn−O(1) (Å) | 1.937 | - | - | - |
| Mn−O(2) (Å) | 2.274<br>1.680 | - | - | - |
| <Mn−O> (Å) | 1.964 | 1.954 | - | 1.958 |
| Mn−O(1)−Mn (º) | 178.5 | - | - | - |
| Mn−O(2)−Mn (º) | 153.8 | - | - | - |
| <Mn−O−Mn> (º) | 162.0 | 165.2 | - | 172.4 |

The increase of the cell volume from LCMO to LBMO is associated with the increase of the mean ionic radius <$r_A$> at the A-site, which results in a reduction of the internal chemical pressure in the perovskite unit cell [19]. This leads to a reduction of the distortion in the MnO$_6$ octahedra, and a consequent structural transition from $Pnma$ to $R\bar{3}c$. Careful examination of the structural characteristics of the samples revealed that while the Mn−O distances along the different octahedral axes vary significantly in LCMO, these distances remained constant in both LSMO and LBMO. Also, examination of the angles between the octahedral axes revealed that these angles in LCMO are ~ 5º, 9º, and 11º different from the ideal value of 90º for a perfect octahedron. However, the angles within LSMO octahedra are only ~ 0.8º off, and in LBMO the angles are < 0.2º off the ideal value of 90º. These results confirm the reduction of the octahedral distortions in going from LCMO to LBMO. Further, the <Mn−O−Mn> angle increases toward the ideal 180º value in going from LCMO to LBMO (see the values given in Table 1), indicating reduction of the structural distortion and improvement of the long-range structural coherence.

The crystallite size for the perovskite phase was determined using Scherrer equation [49]:

$$D = \frac{k\lambda}{\beta \cos(\theta)} \quad (1)$$

Here, $k = 0.91$ for a Gaussian peak profile, $\lambda$ (= 1.54 Å) is the X-ray radiation wave length, and $\beta$ is the full width at half maximum of the diffraction peak. The main structural peaks at $2\theta$ = 32.788º for LCMO, 32.69º for LSMO, and 32.41º for LBMO samples were fitted with Gaussian line shape, from which the peak position and full width at half maximum were determined. The peak width was then corrected for instrumental broadening using a standard silicon sample, and the corrected peak width was used to evaluate the mean crystallite size in each sample. The analysis indicated that the mean cystallite size for LCMO, LSMO, LBMO1000, and LBMO1100 were 20 nm, 23 nm, 31 nm, and 44 nm, respectively. These results indicated that significat improvement of the crytallization of the perovskite phase is achieved by Ba substitution for La, and the increase of the sintering temperature.

### 3.2. SEM measurements

SEM images of all samples indicated that the samples are mainly composed of aggregates of submicron, irregularly-shaped particles (Fig. 3). The images of all samples sintered at 1000º C (LCMO (a); LBMO 1000 (b), and LSMO (d)) revealed that the majority of the samples consisted of particles in the size range of ~ 100 – 300 nm, and that particles with larger size are present with a smaller fraction. Although differences between the images are not obvious, careful examination of the particle size in the various samples makes us believe that the mean particle size in LBMO 1100 sample is higher than in the samples sintered at 1000º C. This is consistent with XRD data which revealed that LBMO 1100 had the highest crystallite size among all samples. In this sample, the higher sintering temperature is responsible for the improvement of crystallization of the perovskite phase.

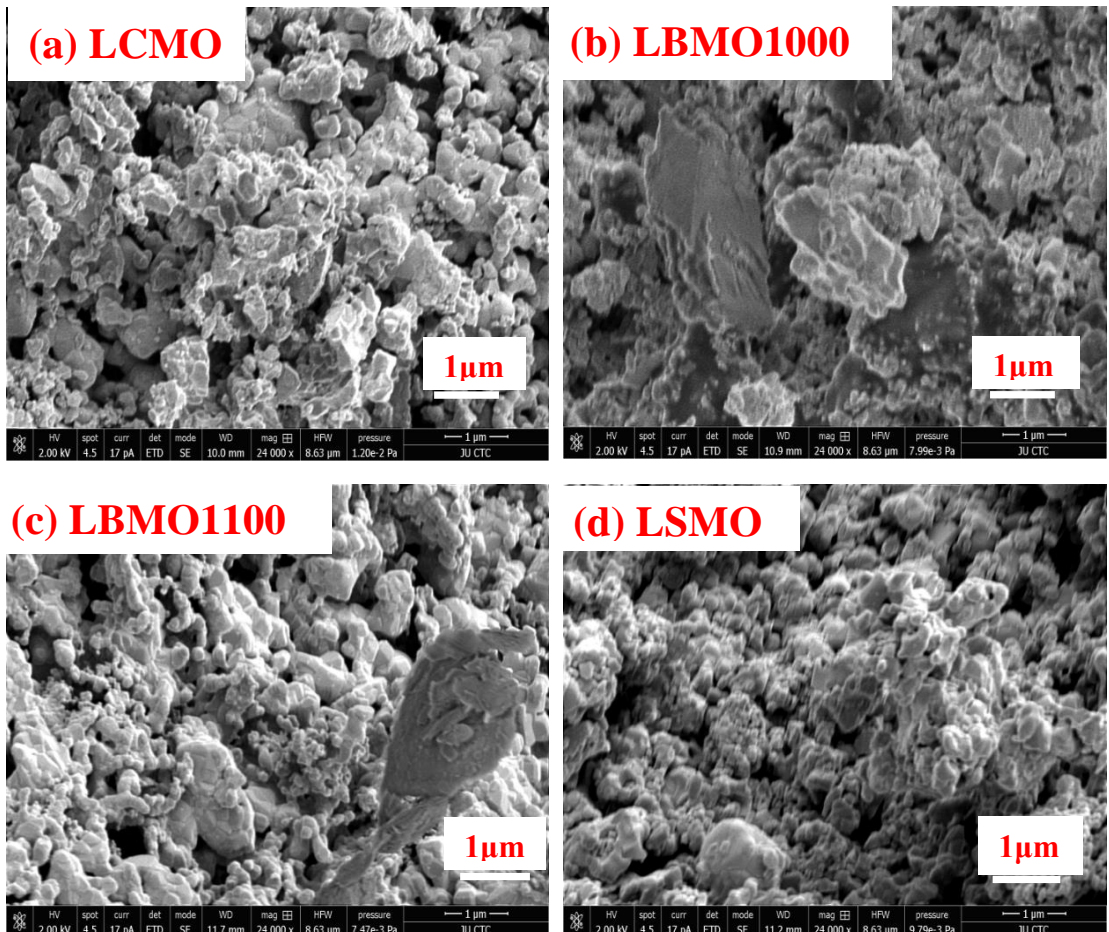

Fig.3: SEM images for (a) LCMO, (b, c) LBMO, and (d) LSMO samples prepared by ball milling

### 3.3. Hysteresis loop measurements

Room temperature hysteresis loops of all samples were recorded at room temperature using VSM under an applied field up to 10 kOe, and are shown in Fig. 4. The hysteresis loops indicated that LCMO is paramagnetic at room temperature, while LSMO and LBMO samples revealed ferromagnetic behavior with low coercivity as shown in the expanded view (Fig. 5). The saturation magnetization of LBMO sample sintered at 1000º C is significantly lower than that of LSMO, which could be associated with the presence of a nonmagnetic impurity phase in LBMO 1000. The saturation magnetization of LBMO sintered at 1100º C, however, exhibited a significant increase, approaching the saturation magnetization of LSMO as demonstrated by Fig. 4.

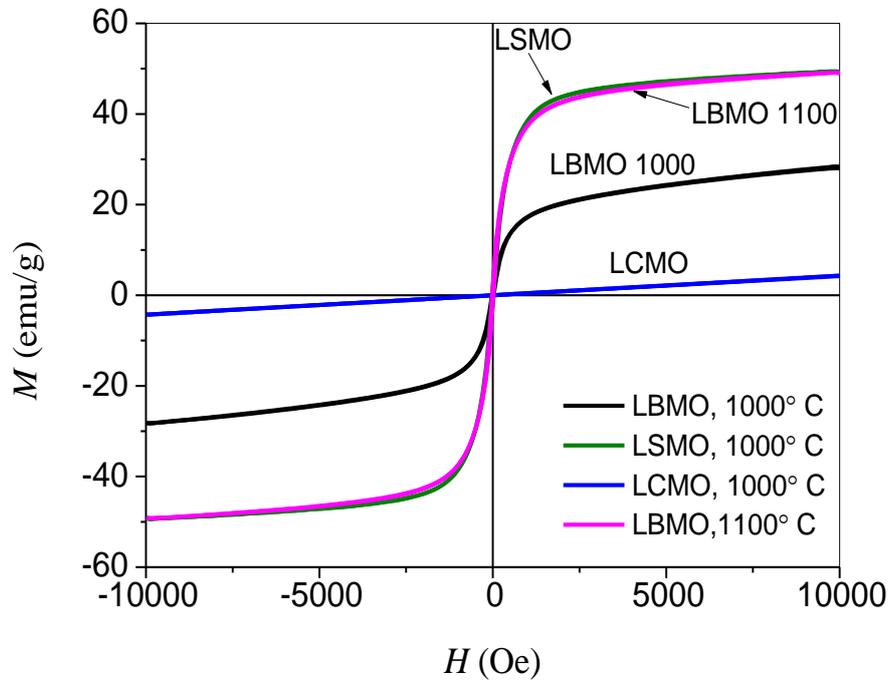

Fig. 4: Hysteresis loops for samples ball milled samples sintered at 1000˚ C.

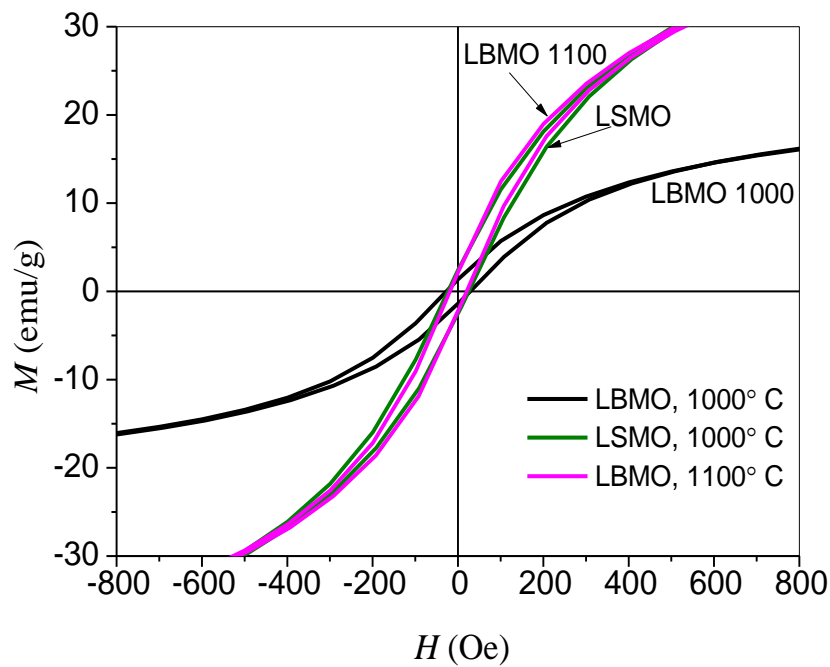

Fig. 5: Expanded hysteresis loops for LBMO and LSMO samples.

Low values of remnant magnetization and coercive fields ($H_c$ < 20 Oe) characteristic of soft magnetic materials for LSMO and LBMO samples were determined directly from the hysteresis loops, and the results are listed in Table 2. However, accurate determination of the saturation magnetization can be achieved by applying the law of approach to saturation in the high field range. In this range, the rotation of domain magnetization is dominant, where the magnetization is given by [50]:

$$M = M_s \left(1 - \frac{A}{H} - \frac{B}{H^2}\right) + \chi H \qquad (2)$$

Here $M_s$ is the spontaneous saturation magnetization of the domains per unit volume, $A$ is a constant representing the contributions of inclusions/microstress, $B$ is a constant representing the contribution of magnetocrystalline anisotropy, and $\chi H$ is the forced magnetization term. In a uniaxial crystal, the contribution of the magnetocrystalline anisotropy is completely determined by the first anisotropy constant $K_1$, where the constant $B$ for a cubic crystal is given by [51]:

$$B = \frac{0.07619\, K_1^{\,2}}{M_s^2} \qquad (3)$$

Accordingly, the law of approach to saturation was used to obtain the saturation magnetization and magnetocrystalline anisotropy of LSMO and LBMO samples. A plot of $M_s$ vs. $1/H^2$ in the high field region (8.5 kOe < $H$ < 10 kOe) gave a straight line, indicating that the magnetocrystalline term is dominant in this field range. The saturation magnetization ($M_s$) was determined from the intercept of the straight line with the magnetization axis, whereas the constant $B$ was determined from the slope, from which the first anisotropy constant was determined using Eq. 3. The derived magnetic parameters of LBMO and LSMO samples are listed in Table 2.

**Table 2:** Saturation magnetization ($M_s$), remanence ($M_r$), coercive field ($H_c$), and first anisotropy constant ($K_1$) for the samples.

| Sample | $M_s$ (emu/g) | $M_r$ (emu/g) | $H_c$ (Oe) | $K_1$ ($10^5$ erg/g) |
|---|---|---|---|---|
| LBMO 1000 | 30.7 | 1.03 | 18.9 | 3.2 |
| LBMO 1100 | 50.8 | 1.43 | 15.1 | 3.2 |
| LSMO | 50.9 | 1.63 | 18.6 | 3.1 |

The saturation magnetization of LBMO increased from 30.7 emu/g for the sample sintered at 1000º C, to 50.8 emu/g as the sintering temperature increased to 1100º C.

This significant improvement of the saturation magnetization is a result of the reduction of the nonmagnetic phase, and improvement of the crystallization of the perovskite phase as a consequence of the increase of the sintering temperature. The saturation magnetization of LBMO 1100 is almost equal to that of LSMO, which is in good agreement with the value of 52 emu/g reported for $La_{0.67}Ba_{0.22}Sr_{0.11}MnO_3$ prepared by sol – gel and sintering at 1100º C [2]. However, the observed saturation magnetization is higher than the value of ~ 47 emu/g reported for $La_{0.7}Sr_{0.3}MnO_3$ prepared by coprecipitation and sintering at 900º C [52]. As for the magnetocrystalline anisotropy, the first anisotropy constant was found to be almost the same for LSMO and LBMO, indicating similar effect for Ba and Sr on the structural and magnetic properties of the perovskite manganites. This conclusion is also consistent with the similarity of the saturation magnetization of the pure LSMO and LBMO perovskites.

### 3.4. Thermomagnetic measurements

The thermomagnetic curves at constants applied field of 100 Oe were measured for all samples sintered at 1000º C. A homogeneous sample with sharp magnetic phase-transition at the Curie temperature ($T_c$) should give a sharp peak in the derivative of the thermomagnetic curve at that temperature. The magnetization curve of LCMO (Fig. 6) exhibited monotonic decrease with the increase of temperature, with no peaks in the derivative curve, indicating that the sample does not exhibit ferromagnetic behavior above room temperature. The derivative curves for LSMO and LBMO samples, however, exhibited peaks corresponding to magnetic phase transitions above room temperature (see Fig. 7).

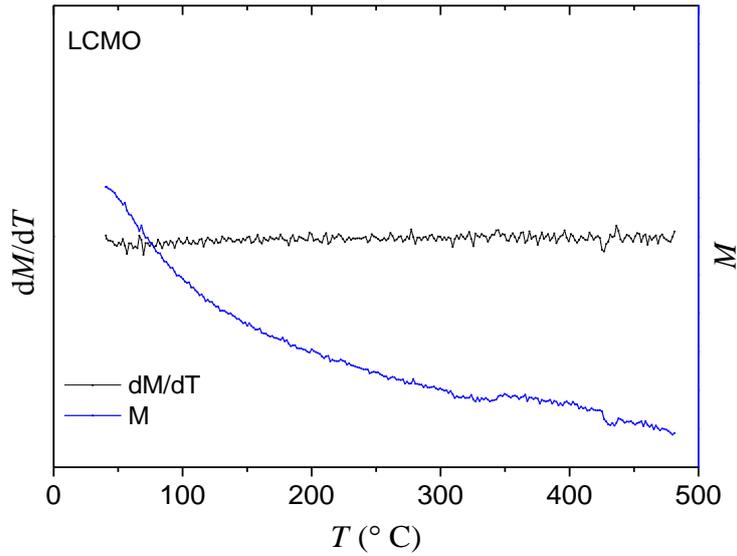

Fig 6: Thermomagnetic curve (at an applied field of 100 Oe) and its derivative for LCMO sample.

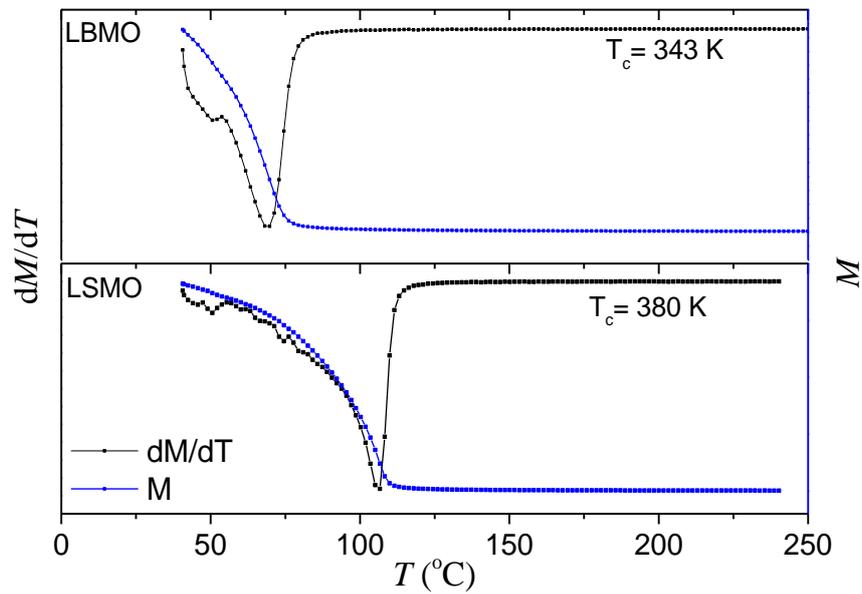

Fig 7: Thermomagnetic curves (at an applied field of 100 Oe) and their derivatives for LSMO and LBMO samples.

The derivative curve of LSMO sample showed a single peak at 380 K, corresponding to ferromagnetic – paramagnetic phase transition with Curie temperature somewhat higher than previously reported values (see data in Table 3). The broadening of the magnetic transition, however, could be evidence of a small degree of magnetic inhomogeneity arising from small local structural distortions in the sample. On the other hand, the derivative curve of LBMO sample showed a double peak structure, which could be an indication of magnetic phase separation, and the presence of a minority magnetic phase, in addition to the major perovskite phase. The temperature (343 K) at which the major peak occurred in the derivative curve is in good agreement with reported values of Curie temperature for LBMO, and this peak is therefore associated with ferromagnetic to paramagnetic phase transition of the major magnetic phase in this sample.

**Table 3**: Curie temperatures of LSMO and LBMO compounds compared to values found in the literature.

| Sample | $T_c$ (observed) | $T_c$ (from literature) |
|---|---|---|
| $La_{0.67}Sr_{0.33}MnO_3$ | 380 K | 342.6 K [19], 370 K [13], 375 K [33], 378 K [1] |
| $La_{0.67}Ba_{0.33}MnO_3$ | 343 K | 335.1 K [19], 332 K [21], 342 K [47] |

## 4. Conclusions

Pure perovskite $La_{0.67}Ca_{0.33}MnO_3$ and $La_{0.67}Sr_{0.33}MnO_3$ phases were synthesized by high energy ball milling and sintering at 1000º C. This synthesis route, however, was not appropriate to produce a pure $La_{0.67}Ba_{0.33}MnO_3$ phase, where a secondary $BaMnO_3$ phase appeared in addition to the major perovskite phase. Single LBMO perovskite phase, however, was obtained by sintering the compound at 1100º C, and a significant increase of the saturation magnetization from 30.7 emu/g to 50.8 emu/g was observed. The saturation magnetization of the sample sintered at 1100º C was almost equal to that of LSMO, and low coercivity of $H_c < 20$ Oe for all samples was observed. The first magnetic anisotropy constant ($K_1$) was found to be almost the same for LSMO and LBMO samples. Room temperature isothermal magnetic measurements, however, revealed that LCMO sample was paramagnetic at room temperature. The thermomagnetic measurements showed ferromagnetic–paramagnetic phase transitions for LBMO and LSMO at 343 K and 380 K, respectively, and revealed evidence of magnetic inhomogeneity in both samples.